\documentclass{Interspeech}

\interspeechcameraready

\usepackage{listings}
\usepackage{pgfplots}
\usepackage{multirow}

\usepackage{float}  
\usepackage{pgfplotstable}
\usepackage{colortbl} 
\usepackage{array} 
\usepackage{dashrule}
\usepackage{cite}
\usepackage{colortbl}
\usepackage{pgfplots}
\usepackage{makecell}
\usepackage{subfig}

\usepackage{url}
\usepackage{graphicx}
\pgfplotsset{compat=1.18}
\definecolor{myblue}{HTML}{DCE6F1} 
\definecolor{mydarkblue}{HTML}{305496} 
\usepackage{hyperref}
\usepackage{amsthm}

\title{Multilingual Source Tracing of Speech Deepfakes: A First Benchmark}

\author[affiliation={1,2,*}]{Xi}{Xuan}
\author[affiliation={3,4}]{Yang}{Xiao}
\author[affiliation={4}]{Rohan Kumar}{Das}
\author[affiliation={1}]{Tomi}{Kinnunen}

\affiliation{School of Computing}{University of Eastern Finland}{Finland}
\affiliation{Department of Linguistics and Translation}{City University of Hong Kong}{Hong Kong SAR}
\affiliation{School of Computing and Information Systems}{The University of Melbourne}{Australia}
\affiliation{}{Fortemedia Singapore}{Singapore}

\email{*Corresponding author: xi.xuan@uef.fi}
\keywords{Source Tracing, Speech Deepfakes, Cross-Lingual generalization, Linguistic Diversity, Unseen Speaker}

\usepackage{comment}

\begin{document}

\maketitle

\begin{abstract}
   Recent progress in generative AI has made it increasingly easy to create natural-sounding deepfake speech from just a few seconds of audio. While these tools support helpful applications, they also raise serious concerns by making it possible to generate convincing fake speech in many languages. Current research has largely focused on detecting fake speech, but little attention has been given to tracing the source models used to generate it. This paper introduces the first benchmark for multilingual speech deepfake source tracing, covering both mono- and cross-lingual scenarios. We comparatively investigate DSP- and SSL-based modeling; examine how SSL representations fine-tuned on different languages impact cross-lingual generalization performance; and evaluate generalization to unseen languages and speakers. Our findings offer the first comprehensive insights into the challenges of identifying speech generation models when training and inference languages differ. The dataset, protocol and code are available at \url{https://github.com/xuanxixi/Multilingual-Source-Tracing}.
\end{abstract}

\section{Introduction}

Recent advances in Generative AI (GenAI) have resulted in an unprecedented surge in synthetic data generation. The National Security Agency (NSA), Federal Bureau of Investigation (FBI), and Department of Homeland Security (DHS) recently released a joint report\footnote{\url{https://www.dhs.gov/sites/default/files/publications/increasing_threats_of_deepfake_identities_0.pdf}}. It warns that synthetic media, especially deepfake content, is now spreading quickly across many languages worldwide. This warning comes at a time when generative AI has made strong progress. Voice synthesis tools, including text-to-speech (TTS)~\cite{Liu2025VoxpopuliTTS} and voice 
conversion (VC)~\cite{Yao2024PromptVC}, can now create very natural speech from just a few seconds of audio~\cite{Li2025DeepfakeSurvey}. These tools help with positive applications like virtual assistants. However, they can also be used to create harmful deepfake speech in several languages. These harmful use cases include phone scams, disinformation and defaming campaigns, and spoofing voice biometric systems to mention a few~\cite{kolupuri2025scams,zhang1,ding1,zhang3,zhang4}.

\begin{figure}[t]
    \centering
    \includegraphics[width=0.45\textwidth]{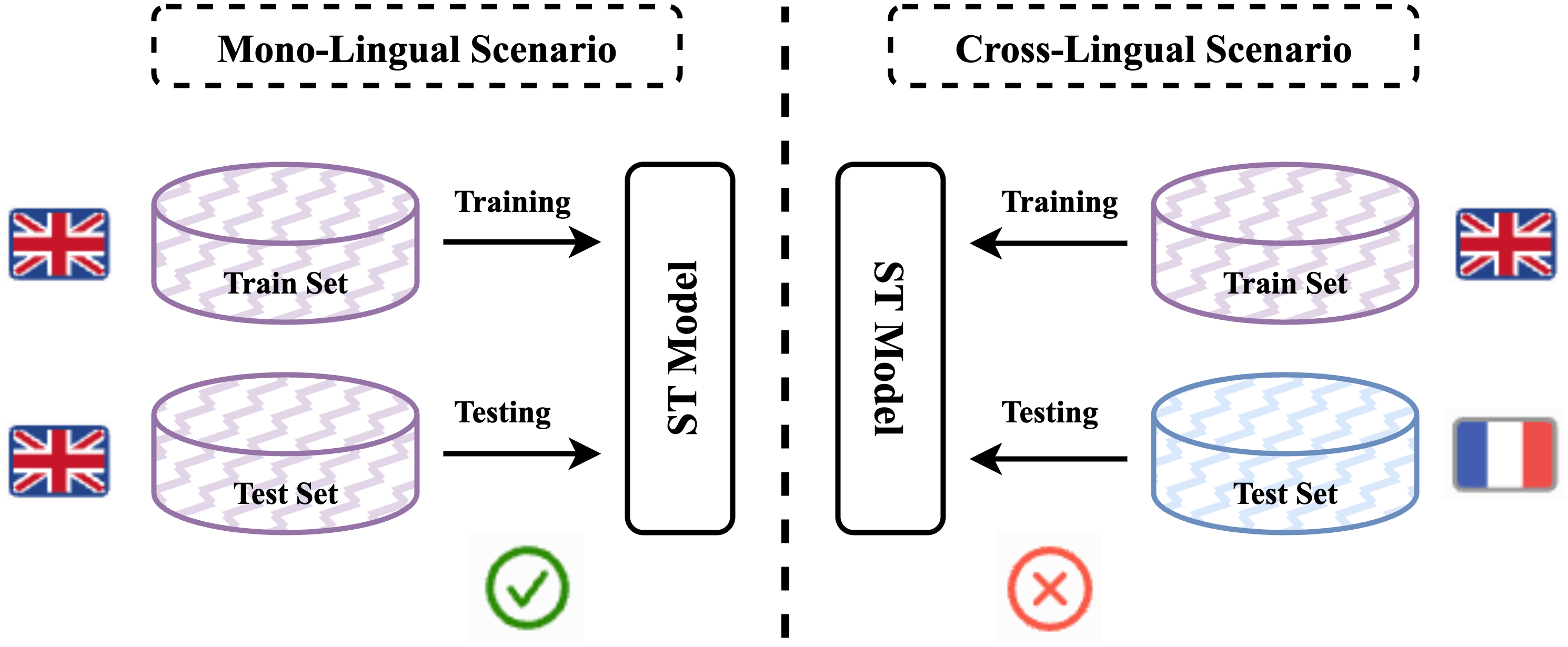}
    \caption{Illustration of mono-lingual and cross-lingual scenarios on source tracing systems. The English data-trained system works well with English data (left) but fails with other languages (right).}
    \label{fig:mono_cross_lingual}
\end{figure}

To address the growing threats of deepfake speech, international challenges like ASVspoof \cite{Liu2023ASVspoof2021, Wang2024ASVspoof5} and audio deepfake
 detection (ADD) \cite{Yi2022ADD2022, Yi2023ADD2023} promote development of new defense methods to detect deepfake speech \cite{Doan2024Trident,ref35,liu2025nes2net}. Some of these detectors are already very accurate, reaching equal error rates (EERs) below 0.5\% on ASVspoof19 \cite{li2024interpretable}. 
But checking whether the audio is real or fake is insufficient; it is imperative to trace the source of a deepfake speech sample---who created it (or, more practically, \emph{which generative architecture is a likely origin of the deepfake speech sample}). In forensic science, source tracing (ST) of evidentiary materials, including telephone recordings~\cite{Faundez2010SpeechWatermarking}, digital images~\cite{Swaminathan2008DigitalImage}, and text documents~\cite{Richter2018ForensicAnalysis}, has been extensively studied for law enforcement applications. Prior research from audio forensics has established methodologies for tracing microphones~\cite{Qamhan2023TransformerAuth,linprime}, cell phones~\cite{Jansen2007CellPhoneForensics,Hanilci2014-source-cellphone-nonspeech}, and caller networks~\cite{Catanese2013ForensicCallNetworks,zhang2}. While audio forensic source analysis is a well-established field, ST for speech deepfakes has emerged as a new and pressing research challenge, receiving thus far relatively little attention.

\begin{figure*}[t]
    \centering
    \includegraphics[width=\textwidth]{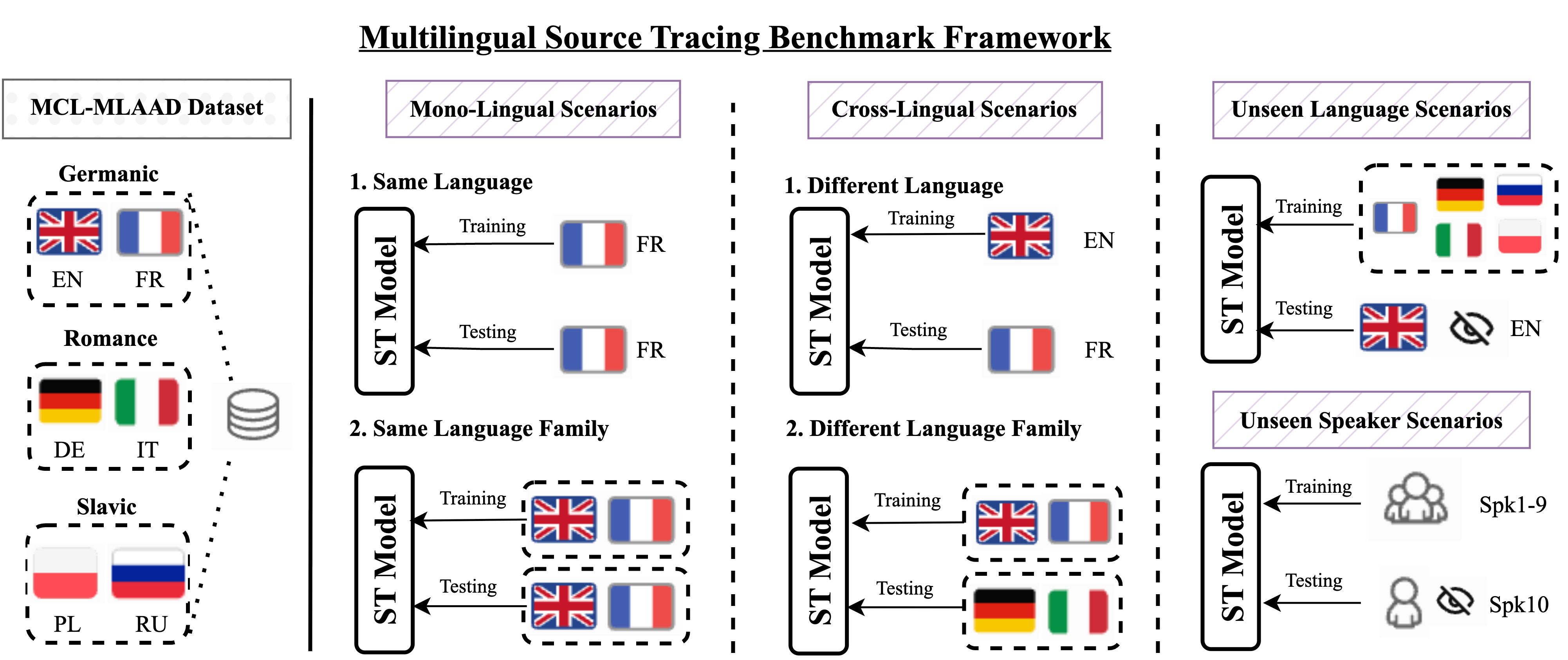}
    \caption{Overview of the multilingual source tracing (ST) benchmark framework. The framework evaluates model generalization across languages and speakers through four experimental scenarios: (1) \textit{Monolingual} (same-language/family training and testing); (2) \textit{Crosslingual} (train on one language/family, test on another language/family); (3) \textit{Unseen languages generalization} (train on multiple languages, test on unseen language); (4) \textit{Unseen speaker generalization} (train on multiple languages and speakers, test on unseen speakers across multiple languages).}
    \label{fig:benchmark}
    \vspace{-2mm}
\end{figure*}

Some early studies on ST give promising results. For example, 
~\cite{klein2024source} compared different ST system designs (end-to-end or two-stages), and 
~\cite{xie2024generalized} proposed a general ST framework which was later extended to a benchmark for neural-codec models~\cite{xie2025neural}. However, these studies are all limited to a single language (specifically, English) for training and testing. They do not study what happens when the model is trained in one language but tested in another. As illustrated in Figure \ref{fig:mono_cross_lingual}, the language mismatch effects refer to the decrease in performance resulting from discrepancies between the languages used during training and inference. These effects have been extensively observed since different languages (and language families) differ at lexical, prosodic, and phonotactic levels – any data-driven model trained on one language only is expected to be overfit to the training language, hindering generalization to new languages. This general problem, called \emph{language mismatch}, is well-known in many tasks including translation~\cite{li2021multilingual, Sin2025, Xuan2025}, speaker anonymization~\cite{meyer2024probing}, and speaker verification~\cite{song2024introducing,xuanasv1, xuanasv2, xuanasv3, xuanasv4, xuan2024conformer}. Its impact on ST, however, remains thus far unknown.
At the same time, audio language models are growing fast. They now support thousands of languages \cite{Pratap2024Scaling}. With only a short voice sample, they can create high-quality speech in any target language \cite{wang2023neural}. This makes it easier to create fake speech that can escape current detection tools. While some new datasets include many languages \cite{hou2024polyglotfake, qi2024madd, muller2024mlaad}, they all focus on the detection task, not tracing which model made them. Our study seeks to fill this knowledge gap.

Recent related studies have advanced multilingual deepfake \emph{detection}. For instance, 
\cite{phukan2024heterogeneity} 
studied how multilingual pre-trained models can detect speech deepfakes across accents and tones, whereas \cite{ba2023transferring} transferred detection knowledge across languages using domain adaptation methods. The study in \cite{liu2024towards}, in turn, proposed accent-based data expansion to reduce the effects of language mismatch \cite{liu2024towards}. Finally, \cite{marek2024audio} 
tested multilingual ability of detection models, demonstrating that training only on English limits their effectiveness in other languages. Despite progressing understanding of the impact of language in the context of deepfakes, all the prior studies share a key limitation: they focus on detection, rather than ST. 

In contrast, our work shifts the focus from binary detection to ST. This task brings new challenges, especially in cross-lingual conditions where the model must trace the origin of fake speech even when the training and testing languages differ. To address this gap, we propose to first explore a benchmark for ST in both mono- and cross-lingual scenarios, and evaluate it across multiple feature types, language settings, and low-resource conditions. 
\textbf{We introduce the first multilingual benchmark for ST of speech deepfakes, covering both mono-lingual and cross-lingual scenarios. Furthermore, this work is the first to explore the effects of unseen languages and speakers in ST tasks}. The speech data itself is drawn from Multi-Language Audio Anti-Spoofing (MLAAD) dataset\cite{muller2024mlaad} used in multiple recent studies, including  studies under Interspeech 2025 special session \emph{Source tracing: The origins of synthetic or manipulated speech}. Despite consisting of an extensive collection of deepfake samples in multiple languages, MLAAD does not contain an evaluation protocol to address ST in both mono-lingual and cross-lingual scenarios. 
Our new benchmark is intended as a reproducible starting point to foster further work in performance assessment and improving multilingual ST models.

Our immediate scientific aim is to better understand how language differences and resource limitations influence ST performance. We address seven key research questions (RQs):

\begin{enumerate}[label=\textbf{RQ\arabic*}, wide=0pt, labelwidth=!, align=left]
\footnotesize 
    \item What is the performance of ST models in monolingual scenarios?
    \item How effectively do ST models generalize to cross-lingual scenarios?
    \item Does training and testing within the same language family improve cross-lingual performance?
    \item How do digital signal processing (DSP) and self-supervised learning (SSL)-based  ST models compare in cross-lingual generalization?
    \item How do training strategies and linguistic diversity of pre-training corpora affect the cross-lingual generalization performance of SSL-based ST models?
    \item How do unseen languages impact ST models generalization?
    \item How do unseen speakers impact ST models robustness against shortcut learning?
\end{enumerate}

By answering these questions, our work offers the first systematic evaluation of multilingual ST and lays a foundation for future research in this area.

\section{Multilingual Source Tracing Benchmark}
As shown in Figure \ref{fig:benchmark}, to comprehensively and fair evaluate Multilingual ST model performance in both \textbf{M}ono- and \textbf{C}ross-\textbf{L}ingual scenarios, we propose the linguistically-balanced dataset \textbf{MCL}-MLAAD and establish novel protocols for thorough assessment of generalization capabilities.

\subsection{Dataset}

Accurate evaluation of Multil-lingual ST models needs a dataset with balanced language and TTS synthesizer distribution. For this, we present the Lingual-Balanced MLAAD dataset. It is an improved version of the original MLAAD corpus~\cite{muller2024mlaad}\footnote{\url{https://deepfake-total.com/sourcetracing}}, which contains 420.7 hours of synthetic speech in 38 languages, produced by 91 TTS models (including 32 types of architectures). This data was created using the M-AILABS multilingual audio source\footnote{\url{https://huggingface.co/datasets/mueller91/MLAAD}}. However, the original corpus has two limitations. First, no single TTS model covers all 38 languages. Second, each language does not include all the 91 TTS models. These limitations hinder multilingual generalization studies.

To address this, we built a refined version of MLAAD v5, referred to as MCL-MLAAD. We select six languages from three language families: Germanic (English, German), Romance (French, Italian), and Slavic (Polish, Russian). We also include four popular TTS architectures: Griffin-Lim, Bark, XTTS v1.1, and XTTS v2. Our dataset design aligns with the partitioning methods in~\cite{marek2024audio}. To simulate diverse acoustic environments, the dataset includes four types of noise perturbations—namely, noise, music, babble, and reverberation from MUSAN \cite{snyder2015musan}—systematically applied to each clean utterance. Thus, five acoustic variants (the clean original and its four perturbed counterparts) are generated per utterance, enhancing real-world robustness. Due to uneven spoofing attack distribution, we partitioned the data into train, dev, and test sets with a 60:20:20 ratio for each language  as detailed in Table~\ref{tab:language_codes}.

\begin{table}[H]
\centering
\small 
\setlength{\tabcolsep}{4pt} 
\caption{Dataset Statistics for Different Languages}
\label{tab:language_codes}
\begin{tabular}{ccccc}
\Xhline{1.2pt}
\textbf{Language} & \textbf{Language} & \textbf{Code} & \textbf{Samples} & \textbf{Dur} \\
  \textbf{Family} & & & \textit{(×5)} & \textit{(×5)} \\
\Xhline{0.8pt}
Germanic & English & en & 2,100 & 4h 37m 59.59s \\
 & German & de & 2,100 & 4h 35m 7.89s \\
Romance & French & fr & 2,100 & 5h 6m 38.35s \\
 & Italian & it & 2,100 & 4h 14m 4.82s \\
Slavic & Polish & pl & 2,100 & 5h 12m 4.68s \\
 & Russian & ru & 1,200 & 3h 11m 16.03s \\
\Xhline{0.8pt}
\textbf{Total} & - & - & \textbf{11,700} & \textbf{26h 57m 11s}\\
\Xhline{1.2pt}
\end{tabular}
\end{table}

\subsection{Protocols}

Our benchmark evaluates source tracing capabilities through four experimental protocols designed to isolate critical dimensions of multilingual generalization:

\subsubsection{Mono- \& Cross-Lingual Protocol}
We trained ST models using only one language each: English, German, French, Italian, Polish, or Russian. These models help us study how the model performs within a single language. Later, we also use them for cross-lingual experiments by testing them on different languages.

\subsubsection{Mono- \& Cross Language Family Protocol}
To explore how language families affect performance, we grouped the six languages into three families: Germanic, Romance, and Slavic. We then trained one model per family. These models help examine whether training on one language group improves performance on related languages.

\subsubsection{Seen \& Unseen Languages Protocol} To investigate the generalization capability for unseen languages, we employ a leave-one-language-out experimental protocol. Our work explicitly contrasts "seen" and "unseen" language conditions and rigorously analyzes how pre-training data affects cross-lingual robustness.

\subsubsection{Seen \& Unseen Speakers Protocol} Besides content-related variability (which manifests as phonemic differences between languages), speaker-related factors can be expected to impact ST performance. For instance, if training data for TTS method $A$ represents only one voice identity $1$ while training data for TTS method $B$ represents only voice identity $2$, a model may learn to differentiate the two speakers, as opposed to the two sources---an instance of \emph{shortcut learning} \cite{geirhos2020shortcut}. 
To analyze the impact of seen vs. unseen speakers, we use a leave-one-language-out protocol: the model is trained on all but one language, which serves as an unseen test language.

This analysis brings up some new challenges. As opposed to language labels, the MLAAD metadata does not contain speaker labels. Moreover, arguably synthetic speech does not even \emph{have} crisply-defined speaker identity (only targeted speaker identity). These necessitate approximate, \emph{pseudo-speaker} labels that we obtain through an approach similar to \cite[Section 3.3]{klein2024source}. We use an off-the-shelf\footnote{\url{https://huggingface.co/Jenthe/ECAPA2 }} ECAPA2~\cite{thienpondt2023ecapa2} model to extract speaker embeddings from each synthetic utterance, followed by clustering the resulting $11,700$ embeddings using spherical k-means \cite{Dhillon2001-spherical-kmeans}, a method suitable for clustering length-normalized $d=192$ dimensional embeddings. 
We first run 10 repeats (each with different random initialization) of spherical k-means for each cluster count in $k \in [1,100]$ and use an 'elbow' criterion \cite{kodinariya2013review} to set the number of clusters. We then repeat clustering with the selected cluster count ($k=18$) using 100 restarts. The cluster assignments of speaker embeddings give us unique pseudo-speaker label per utterance.

\begin{table}[htbp]
    \centering
    \caption{Protocol statistics for speaker effect analysis.}
    \label{tab:sample_threshold_unified_single_col}
    \renewcommand{\arraystretch}{1.2} 
    \begin{tabular}{cccc} 
        \toprule[1.2pt]
        \textbf{Unseen} & \multicolumn{2}{c}{\textbf{Utterance Count}} & \multirow{2}{*}{\textbf{Threshold}} \\
        \cmidrule(lr){2-3}
        \textbf{Language} & \textbf{Seen spk.} & \textbf{Unseen spk.} & \\
        \midrule[1.0pt]
        English  & 938 & 982 & 0.085 \\
        German   & 1,029 & 891 & 0.075 \\
        French   & 908 & 1,012 & 0.079 \\
        Italian  & 919 & 1,001 & 0.077 \\
        Polish   & 1,026 & 894 & 0.077 \\
        Russian  & 1,064 & 1,036 & 0.074 \\
        \bottomrule[1.2pt]
    \end{tabular}
\end{table}

A pseudo-speaker $i$ is defined as \textit{seen} if 
the empirical speaker prior in the combined training and development data
\( P_i \equiv N_i^{\text{tra+val}} / N_{\text{total}}^{\text{tra+val}} \) exceeds threshold $\theta$, \emph{and} speaker $i$ is present in the test data. Here, \( N_i^{\text{tra+val}} \) and \( N_{\text{total}}^{\text{tra+val}} \) denote speaker-specific and total training-development samples respectively. Here, $\theta$ mitigates the inherent trade-off between strict speaker exclusion (which causes severe data scarcity) and complete inclusion (which induces class imbalance), ensuring statistically viable group comparisons. 
We balanced the number of test samples between the seen and unseen groups. The per-language seen/unseen 
utterance counts and thresholds 
summarized in Table \ref{tab:sample_threshold_unified_single_col}.

\section{Multilingual Source Tracing Methods}

\subsection{ST Models}

\begin{figure}[htbp]
    \centering
    \includegraphics[width=0.7\linewidth]{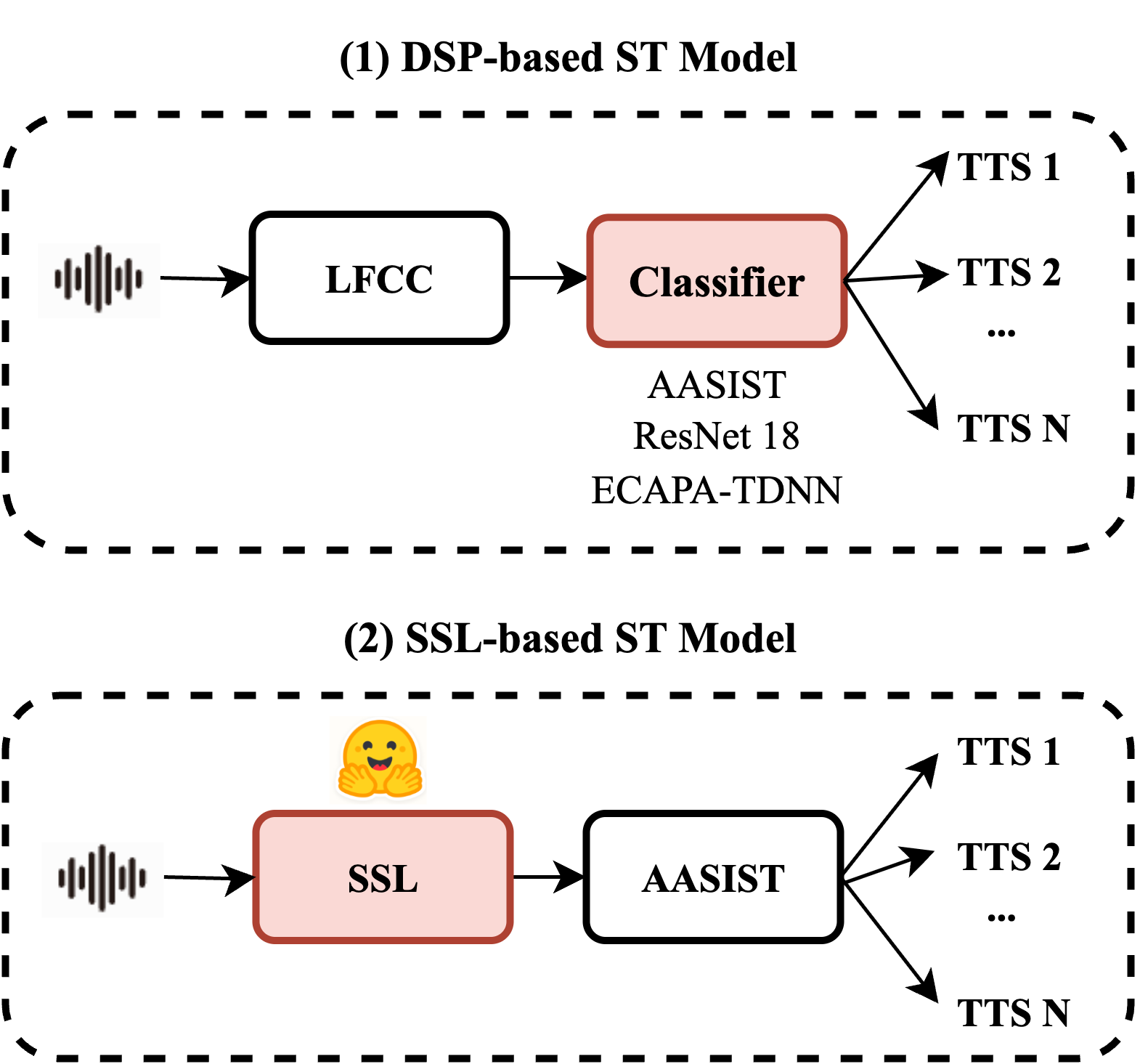}
    \caption{Architecture Comparison of DSP-based and SSL-based Source Tracing (ST) Models. 
        Red components represent variable components that can be replaced or adjusted in different scenarios, 
        while white components represent fixed components that remain consistent across the models.
    }
    \label{fig:st_model_architecture}
\end{figure}

This section introduces the models used for ST tasks. We first study how different input features influence performance in both mono-lingual and cross-lingual settings. Based on the type of front-end feature, we group the models into two categories: DSP-based and SSL-based. The architectural diagrams of both model types are presented in Figure \ref{fig:st_model_architecture}.

\vspace{1ex}
\subsubsection{Models with DSP front-end} 
We developed three models using classic digital signal processing features: LFCC-ResNet18, LFCC-AASIST and LFCC-ECAPA-TDNN. LFCCs (linear frequency cepstral coefficients) 
were used in early synthetic speech detectors \cite{sahidullah2015comparison}. We 
consider three different backends: (1) AASIST \cite{jung2022aasist} is an audio anti-spoofing method using integrated spectro-temporal graph attention networks; (2) ResNet18 \cite{he2016deep} is designed to learn local spectral patterns using residual blocks; (3) ECAPA-TDNN \cite{desplanques2020ecapa} focuses on capturing global information using attention mechanisms and multi-scale features.

\vspace{1ex}
\subsubsection{Models with SSL front-end} We also develop eight models that use self-supervised learning (SSL) front-ends with a shared back-end, AASIST \cite{jung2022aasist}. These front-ends include: (1) two foundation models—XLS-R-300M \cite{conneau2021unsupervised}, trained on multilingual data \cite{wang2021voxpopuli, pratap2020mls, ardila2019commonvoice, valk2021voxlingua, gales2014babel}, and wav2vec2.0 Large LV-60 \cite{baevski2020wav2vec}, trained on English \cite{panayotov2015librispeech}; (2) six versions of XLS-R fine-tuned on specific languages. The XLS-R builds on wav2vec2.0 by enabling cross-lingual learning. It does this through shared quantization over encoded features, allowing different languages to share acoustic representations \cite{pascu2024towards}. The front-end details are summarized in Table~\ref{tab:ssl}.

\begin{table}[htbp]
\centering
\caption{Evaluation of self-supervised representations for ST. All SSL models have 300M parameters. Base models include wav2vec2.0 Large LV-60 and XLS-R-300M. Language-specific fine-tuned variants are based on large-xlsr-53, trained on six languages (en, de, fr, it, pl, ru). The datasets abbreviations are: Librispeech (LL)~\cite{panayotov2015librispeech}, CommonVoice (CV)~\cite{ardila2019commonvoice}, BABEL (BBL)~\cite{gales2014babel}, multilingual Librispeech (MLS)~\cite{pratap2020mls}, VoxPopuli (VP)~\cite{wang2021voxpopuli}, and VoxLingua107 (VL)~\cite{valk2021voxlingua}.}
\label{tab:ssl}
\scriptsize 
\setlength{\tabcolsep}{3pt} 
\renewcommand{\arraystretch}{1.0}
\begin{tabular}{@{}lcccp{1.8cm}@{}} 
\toprule[1.2pt]
\textbf{Name} & \multicolumn{2}{c}{\textbf{Pretraining}} & \textbf{Fine-tuning} & \textbf{Datasets} \\
\cmidrule(lr){2-3}
\textbf{Model} & \textbf{Dur. (h)} & \textbf{Langs.} & \textbf{Lang.} & \\ 
\midrule[1.2pt]
\textbf{wav2vec2} \\
1\quad large-lv60 & 53k & en & -- & \tiny LL \\
2\quad xls-r-300m & 436k & many & -- & \tiny CV, BBL, MLS, VP, VL \\
\midrule
\textbf{Fine-tuned variants} \\
3\quad large-xlsr-53-en & 56k & many & en & CV-en \\
4\quad large-xlsr-53-de & 56k & many & de & CV-de \\
5\quad large-xlsr-53-fr & 56k & many & fr & CV-fr \\
6\quad large-xlsr-53-it & 56k & many & it & CV-it \\
7\quad large-xlsr-53-pl & 56k & many & pl & CV-pl \\
8\quad large-xlsr-53-ru & 56k & many & ru & CV-ru \\
\bottomrule[1.2pt]
\end{tabular}
\end{table}

\subsection{Implementation details}

All audio samples were downsampled to 16 kHz and trimmed or padded to 4 seconds (64,000 samples). Multiclass cross-entropy loss was used to train all the models with a batch size of 16, and 50 epochs. We used an initial learning rate of \(5 \times 10^{-4}\), and selected the final model based on the lowest development loss. 
\vspace{1ex}

\subsubsection{Models with DSP front-end}LFCCs are extracted using 20ms window and 10ms shift, producing 
feature matrices of shape (80, 399), where 80 is the number of LFCCs and 399 is the number of frames. We implemented two classifiers: ResNet18\footnote{\url{https://github.com/hubert10/ResNet18_from_Scratch_using_PyTorch/blob/main/resnet18.py}} and ECAPA-TDNN\footnote{\url{https://github.com/TaoRuijie/ECAPA-TDNN/blob/main/model.py}}. 

\vspace{1ex}
\subsubsection{Models with SSL front-end} 
 
Each SSL front-end converts raw audio into a matrix of size $199 \times 1024$, where $199$ is the time frame and 1024 is the feature dimension. These were linearly projected to 128 dimensions before being passed to the AASIST classifier\footnote{\url{https://github.com/clovaai/aasist/blob/main/models/AASIST.py}}. During training, we used Mixup \cite{zhang2018mixup, cui2025unlocking}, a method that blends two samples and their labels. The mixing ratio $\lambda$ is drawn from a Beta distribution with parameters $\alpha = 0.5$. This process keeps the input size unchanged and adds a label smoothing effect, which helps improve model generalization.

\vspace{1ex}
\subsubsection{Metrics} 
Experimental results were quantified using (Macro-averaged) Macro-F1 metric: 
Macro-F1 is defined as 
$$
\text{Macro-F1} = \frac{2 \cdot \overline{P} \cdot \overline{R}}{\overline{P} + \overline{R}},
$$
where $\overline{P}$ and $\overline{R}$ denote the average precision and recall across all classes (synthesizer). 
Precision and recall are computed per-class as 
$$
P = \frac{TP}{TP + FP} \quad \text{and} \quad 
R = \frac{TP}{TP + FN},
$$
respectively, with $\mathit{TP}$, $\mathit{FP}$, and $\mathit{FN}$ representing true positives, false positives, and false negatives. 
This aligns with evaluation metrics from recent studies \cite{klein2024source}.

\section{Results and Discussion}

\subsection{Mono-Lingual Performance (RQ1)}
As shown in the diagonal entries of Table~\ref{tab:compar_dsp_ssl}, monolingual performance demonstrates that W2V2({\it{xx}})-AASIST achieves highest macro-F1 score of 97.91\%, indicating that language-specific fine-tuning enhance phonetic differentiation. Notably, LFCC-ECAPA-TDNN attains 97.78\% (18.98\% higher than LFCC-AASIST), indicating its back-end could more effectively capture subtle artifacts.

\subsection{Cross-Lingual Transfer Performance (RQ2)}
As shown in the off-diagonal entries of Table~\ref{tab:compar_dsp_ssl}, LFCC-ECAPA-TDNN achieved optimal cross-lingual performance (88.40\%), exceeding LFCC-AASIST by 33.94\%, suggesting its back-end better handle phonological variations. $\text{W2V2EN}$-AASIST showed strong English→other transfers (average 95.76\%) but lower performance for non-English pairs, reflecting English-only pretrained SSL bias. Low-resource language transfers exhibit significant performance degradation compared to high-resource pairs, highlighting persistent challenges in modeling typologically distant languages.

\subsection{Language Family Effects (RQ3)}

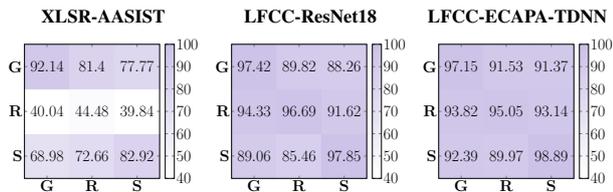
\begin{figure}[htbp]
\centering

\begin{minipage}[t]{0.15\textwidth}
\centering
{\scriptsize \textbf{XLSR-AASIST}}\\[0.5em]
\resizebox{\textwidth}{!}{%
\begin{tikzpicture}
\begin{axis}[
    width=9cm,
    height=9cm,
    xtick={0,1,2},
    ytick={0,1,2},
    xticklabels={$\mathbf{G}$, $\mathbf{R}$, $\mathbf{S}$},
    yticklabels={$\mathbf{G}$, $\mathbf{R}$, $\mathbf{S}$},
    enlargelimits=false,
    axis equal image,
    colormap={custom}{rgb255=(255,255,255) rgb255=(203,192,231)},
    colorbar,
    nodes near coords,
    nodes near coords align={center},
    point meta min=40,
    point meta max=100,
    nodes near coords style={font=\fontsize{25}{24}\selectfont\bfseries},
    tick label style={font=\Huge\bfseries\color{black}},
    xlabel style={at={(axis description cs:0.5,-0.1)}, anchor=north},
    ylabel style={at={(axis description cs:-0.1,0.5)}, anchor=south, rotate=90},
]
\addplot[
    matrix plot,
    mesh/cols=3,
    point meta=explicit
] coordinates {
(0,2)[68.98] (1,2)[72.66] (2,2)[82.92]
(0,1)[40.04] (1,1)[44.48] (2,1)[39.84]
(0,0)[92.14] (1,0)[81.40] (2,0)[77.77]
};
\end{axis}
\end{tikzpicture}}
\end{minipage}
\hfill
\begin{minipage}[t]{0.15\textwidth}
\centering
{\scriptsize \textbf{LFCC-ResNet18}}\\[0.5em]
\resizebox{\textwidth}{!}{%
\begin{tikzpicture}
\begin{axis}[
    width=9cm,
    height=9cm,
    xtick={0,1,2},
    ytick={0,1,2},
    xticklabels={$\mathbf{G}$, $\mathbf{R}$, $\mathbf{S}$},
    yticklabels={$\mathbf{G}$, $\mathbf{R}$, $\mathbf{S}$},
    enlargelimits=false,
    axis equal image,
    colormap={custom}{rgb255=(255,255,255) rgb255=(203,192,231)},
    colorbar,
    nodes near coords,
    nodes near coords align={center},
    point meta min=40,
    point meta max=100,
    nodes near coords style={font=\fontsize{25}{24}\selectfont\bfseries},
    tick label style={font=\Huge\bfseries\color{black}},
    xlabel style={at={(axis description cs:0.5,-0.1)}, anchor=north},
    ylabel style={at={(axis description cs:-0.1,0.5)}, anchor=south, rotate=90},
]
\addplot[
    matrix plot,
    mesh/cols=3,
    point meta=explicit
] coordinates {
(0,2)[89.06] (1,2)[85.46] (2,2)[97.85]
(0,1)[94.33] (1,1)[96.69] (2,1)[91.62]
(0,0)[97.42] (1,0)[89.82] (2,0)[88.26]
};
\end{axis}
\end{tikzpicture}}
\end{minipage}
\hfill
\begin{minipage}[t]{0.15\textwidth}
\centering
{\scriptsize \textbf{LFCC-ECAPA-TDNN}}\\[0.5em]
\resizebox{\textwidth}{!}{%
\begin{tikzpicture}
\begin{axis}[
    width=9cm,
    height=9cm,
    xtick={0,1,2},
    ytick={0,1,2},
    xticklabels={$\mathbf{G}$, $\mathbf{R}$, $\mathbf{S}$},
    yticklabels={$\mathbf{G}$, $\mathbf{R}$, $\mathbf{S}$},
    enlargelimits=false,
    axis equal image,
    colormap={custom}{rgb255=(255,255,255) rgb255=(203,192,231)},
    colorbar,
    nodes near coords,
    nodes near coords align={center},
    point meta min=40,
    point meta max=100,
    nodes near coords style={font=\fontsize{25}{24}\selectfont\bfseries},
    tick label style={font=\Huge\bfseries\color{black}},
    xlabel style={at={(axis description cs:0.5,-0.1)}, anchor=north},
    ylabel style={at={(axis description cs:-0.1,0.5)}, anchor=south, rotate=90},
]
\addplot[
    matrix plot,
    mesh/cols=3,
    point meta=explicit
] coordinates {
(0,2)[92.39] (1,2)[89.97] (2,2)[98.89]
(0,1)[93.82] (1,1)[95.05] (2,1)[93.14]
(0,0)[97.15] (1,0)[91.53] (2,0)[91.37]
};
\end{axis}
\end{tikzpicture}}
\end{minipage}

\caption{
Heatmaps illustrating the Macro F1-score (\%) comparison across XLSR-AASIST, LFCC-ResNet18, and LFCC-ECAPA-TDNN models. Rows indicate source language families used for training (\textbf{G: Germanic}, \textbf{R: Romance}, \textbf{S: Slavic}). Columns indicate target language families used for evaluation.
}
\label{fig:heatmap}
\vspace{-4mm}
\end{figure}

As shown in Figure \ref{fig:heatmap}, cross-family performance analysis reveals consistent advantages for monofamily transfers compared to cross-family settings across all architectures. DSP models demonstrate superior robustness, maintaining minimal performance variance between language pairs while achieving strong cross-family generalization. In contrast, XLSR-AASIST exhibits significant performance degradation under cross-family conditions despite comparable effectiveness in monofamily scenarios, highlighting its heightened sensitivity to linguistic distance. Notably, DSP models preserve near-optimal performance for typologically distant language groups, whereas XLSR-AASIST shows pronounced disparities, suggesting fundamental differences in cross-linguistic divergence modeling.

\begin{table}[htbp]
\centering
\renewcommand{\arraystretch}{0.92} 
\caption{Performance comparison (macro F1-scores, \%) of SSL-based (top three) and DSP-based (bottom three) models across six languages, with the highest macro F1-scores for monolingual (mono) and cross-lingual (cross) settings underlined. SSL-based models: XLSR denotes XLS-R-300M (multilingual pretrained), W2V2EN denotes wav2vec2.0 Large LV-60 (English-only pretrained), and W2V2 (xx) denotes XLS-R (language-specific fine-tuned, language code: xx). Cells show source \textbf{(S)}→target \textbf{(T)} language performance, with diagonals indicating monolingual results and off-diagonals cross-lingual transfer.}
\label{tab:compar_dsp_ssl} 
\captionsetup{font=small,labelfont=bf}  

\begin{tabular}{@{}l|c|c|c|c|c|c@{}}
\toprule[1.5pt]
{\scriptsize S\textbackslash T} & en & de & fr & it & pl & ru \\
\midrule[1.5pt]
\multicolumn{7}{@{}c}{\textbf{XLSR-AASIST} \quad (mono:80.05; cross:59.75)} \\
\midrule
en & \textbf{93.55} & 91.89 & 78.09 & 73.21 & 81.11 & 64.95 \\
de & 80.46 & \textbf{88.53} & 74.78 & 75.17 & 83.28 & 50.11 \\
fr & 33.94 & 35.78 & \textbf{41.43} & 36.95 & 41.19 & 25.42 \\
it & 76.00 & 63.55 & 75.17 & \textbf{82.08} & 70.05 & 59.25 \\
pl & 51.36 & 53.12 & 68.31 & 52.23 & \textbf{92.64} & 39.45 \\
ru & 54.10 & 46.26 & 47.42 & 53.98 & 55.82 & \textbf{82.07} \\
\bottomrule[1.5pt]
\end{tabular}

\vspace{1em} 

\begin{tabular}{@{}l|c|c|c|c|c|c@{}}
\toprule[1.5pt]
\multicolumn{7}{c}{\textbf{W2V2EN-AASIST} \quad (mono:92.84; cross:78.18)} \\
\midrule
en & \textbf{99.51} & 98.64 & 95.06 & 93.09 & 96.61 & 91.63 \\
de & 90.95 & \textbf{66.21} & 89.76 & 64.75 & 67.03 & 71.47 \\
fr & 93.68 & 92.74 & \textbf{98.62} & 90.40 & 98.42 & 90.17 \\
it & 96.64 & 96.13 & 93.75 & \textbf{97.35} & 97.17 & 94.09 \\
pl & 50.08 & 60.70 & 76.30 & 57.39 & \textbf{98.93} & 57.62 \\
ru & 43.59 & 36.72 & 50.28 & 61.04 & 49.45 & \textbf{96.40} \\
\bottomrule[1.5pt]
\end{tabular}

\vspace{1em}

\begin{tabular}{@{}l|c|c|c|c|c|c@{}}
\toprule[1.5pt]
\multicolumn{7}{c}{\textbf{W2V2 ({\it{xx}})-AASIST} \quad (\underline{mono:97.91}; cross:73.45)} \\
\midrule
en & \textbf{99.12} & 97.85 & 93.66 & 95.22 & 95.76 & 91.06 \\
de & 92.63 & \textbf{96.37} & 90.93 & 83.58 & 95.03 & 70.95 \\
fr & 88.29 & 91.91 & \textbf{97.36} & 87.57 & 97.03 & 88.12 \\
it & 94.45 & 88.27 & 92.53 & \textbf{97.53} & 94.85 & 89.33 \\
pl & 43.90 & 54.43 & 66.69 & 44.78 & \textbf{99.16} & 35.31 \\
ru & 47.35 & 37.63 & 59.06 & 60.85 & 62.40 & \textbf{97.91} \\
\bottomrule[1.5pt]
\end{tabular}

\vspace{1em}

\begin{tabular}{@{}l|c|c|c|c|c|c@{}}
\toprule[1.5pt]
\multicolumn{7}{c}{\textbf{LFCC-AASIST} \quad (mono:78.80; cross:54.46)} \\
\midrule
en & \textbf{82.58} & 62.71 & 57.85 & 59.53 & 64.49 & 54.28 \\
de & 56.51 & \textbf{74.33} & 60.50 & 57.15 & 72.57 & 49.25 \\
fr & 58.35 & 62.12 & \textbf{78.40} & 62.28 & 73.17 & 54.79 \\
it & 65.48 & 59.42 & 55.90 & \textbf{73.56} & 65.58 & 58.48 \\
pl & 42.53 & 54.35 & 55.63 & 49.01 & \textbf{87.18} & 43.84 \\
ru & 32.30 & 30.26 & 32.48 & 37.87 & 45.25 & \textbf{76.76} \\
\bottomrule[1.5pt]
\end{tabular}

\vspace{1em}

\begin{tabular}{@{}l|c|c|c|c|c|c@{}}
\toprule[1.5pt]
\multicolumn{7}{c}{\textbf{LFCC-ResNet18} \quad (mono:95.76; cross:79.23)} \\
\midrule
en & \textbf{97.32} & 87.00 & 79.49 & 81.13 & 88.92 & 64.04 \\
de & 93.60 & \textbf{96.44} & 89.99 & 85.52 & 93.78 & 71.86 \\
fr & 90.32 & 87.43 & \textbf{94.86} & 84.22 & 90.34 & 72.08 \\
it & 91.98 & 87.36 & 86.29 & \textbf{92.74} & 91.12 & 82.79 \\
pl & 80.53 & 84.56 & 79.76 & 70.83 & \textbf{97.37} & 55.33 \\
ru & 65.91 & 56.71 & 56.73 & 62.84 & 64.34 & \textbf{95.82} \\
\bottomrule[1.5pt]
\end{tabular}

\vspace{1em}

\begin{tabular}{@{}l|c|c|c|c|c|c@{}}
\toprule[1.5pt]
\multicolumn{7}{c}{\textbf{LFCC-ECAPA-TDNN} \quad (mono:97.78; \underline{cross:88.40)}} \\
\midrule
en & \textbf{98.42} & 94.67 & 92.44 & 89.82 & 94.87 & 78.13 \\
de & 95.95 & \textbf{98.03} & 95.73 & 93.53 & 96.32 & 89.26 \\
fr & 88.86 & 94.65 & \textbf{96.59} & 89.78 & 94.06 & 92.87 \\
it & 96.45 & 96.91 & 95.63 & \textbf{97.19} & 97.19 & 84.69 \\
pl & 79.26 & 87.46 & 81.04 & 72.73 & \textbf{98.76} & 60.01 \\
ru & 83.73 & 84.66 & 84.45 & 81.43 & 85.48 & \textbf{97.98} \\
\bottomrule[1.5pt]
\end{tabular}

\end{table}

\begin{table*}[htbp]
    \centering
    \caption{Macro-Averaged F1 Scores ($\uparrow$) under Leave-One-Language-Out Setting for Seen and Unseen Languages; Method A: XLSR-AASIST; Method B: LFCC-ResNet18; Method C: LFCC-ECAPA-TDNN.}
    \vspace{-2mm}
    \label{tab:merged_f1}
    \resizebox{\textwidth}{!}{%
    \begin{tabular}{l| *{14}{c}}
        \toprule[1.2pt]
        \multirow{2}{*}{Method} & 
        \multicolumn{7}{c|}{\textbf{Seen Languages}} & 
        \multicolumn{7}{c}{\textbf{Unseen Languages}} \\
        \cmidrule(lr){2-8} \cmidrule(l){9-15}
        & -en & -de & -fr & -it & -pl & -ru & Avg & 
        en & de & fr & it & pl & ru & Avg \\
        \midrule[1.2pt]
        A & 
        54.50 & 72.09 & 95.73 & 50.18 & 82.52 & 48.83 & 67.31 & 
        49.61 & 72.46 & 94.39 & 54.12 & 85.40 & 25.35 & 63.56 \\
        
        B & 
        \textbf{97.75} & \textbf{97.49} & 97.68 & \textbf{98.21} & \textbf{97.75} & 95.75 & \textbf{97.44} & 
        96.80 & \textbf{97.06} & 95.55 & 93.00 & 97.88 & \textbf{99.20} & \textbf{96.58} \\
        
        C & 
        97.62 & 94.99 & \textbf{97.96} & 97.53 & 97.45 & \textbf{98.26} & 97.30 & 
        \textbf{96.82} & 96.75 & \textbf{97.16} & \textbf{94.92} & \textbf{98.62} & 91.91 & 96.03 \\
        \bottomrule[1.2pt]
    \end{tabular}}
\end{table*}
\begin{table*}[htbp]
    \centering
    \caption{Macro-averaged F1 scores ($\uparrow$) for three methods on Unseen Languages, evaluated under Seen and Unseen Speaker conditions. Method A: XLSR-AASIST, Method B: LFCC-ResNet18, Method C: LFCC-ECAPA-TDNN.}
    \vspace{-2mm}
    \label{tab:speaker-effects}
    \resizebox{\textwidth}{!}{%
    \begin{tabular}{l| *{14}{c}} 
        \toprule[1.2pt]
        \multirow{2}{*}{Method} & 
        \multicolumn{7}{c}{\textbf{Seen Speakers}} & 
        \multicolumn{7}{c}{\textbf{Unseen Speakers}} \\
        \cmidrule(lr){2-8} \cmidrule(lr){9-15}
        & en & de & fr & it & pl & ru & Avg & en & de & fr & it & pl & ru & Avg \\
        \midrule[1.2pt]
        A        
        & 54.66 & 52.69 & 96.19 & 40.62 & 84.70 & 48.11 & 62.83 
        & 54.63 & 75.78 & 93.97 & 49.88 & 78.91 & 49.13 & 67.05 \\
        
        B        
        & 98.04 & \textbf{86.76} & 97.64 & \textbf{93.39} & 97.59 & 98.16 & \textbf{95.26} 
        & \textbf{96.84} & \textbf{94.69} & 97.65 & \textbf{97.45} & \textbf{97.01} & \textbf{98.35} & \textbf{97.00} \\
        
        C      
        & \textbf{98.67} & 80.63 & \textbf{98.33} & 90.53 & \textbf{98.23} & \textbf{98.42} & 94.14 
        & 96.82 & 91.73 & \textbf{98.37} & 96.12 & 96.61 & 97.71 & 96.23 \\
        \bottomrule[1.2pt]
    \end{tabular}}
    \vspace{-2mm}
\end{table*}

\subsection{Comparison between DSP and SSL Models (RQ4)}

As shown in the off-diagonal entries of Table~\ref{tab:compar_dsp_ssl}, the SSL-based models (top three) and DSP-based models (bottom three) are displayed their cross-lingual performance. When backend architectures are comparable (top four subtables), SSL models English-only pretrained and language-specific fine-tuning SSL front-end demonstrate competitive cross-lingual performance, suggesting that domain adaptation strategies can mitigate inherent language biases. However, DSP-based methods with robust backend ECAPA-TDNN exhibit superior cross-lingual stability, particularly under low-resource conditions. These findings indicate that while SSL models benefit from language-specific fine-tuning to bridge linguistic gaps, DSP architectures inherently prioritize language-agnostic patterns through their signal processing pipelines, offering a more resilient framework for cross-lingual source tracing when paired with advanced backend designs.

\subsection{Training Strategy Effects (RQ5)}
As shown in Table~\ref{tab:compar_dsp_ssl}, this section compares the SSL-based models (top three) and LFCC-based models (bottom three) under varying training strategies to assess their cross-lingual generalization capabilities. Among SSL models, multilingual SSL pretraining (XLSR-AASIST) exhibits lower monolingual performance and weaker cross-lingual transfer capabilities, particularly in low-resource settings. In contrast, English-only SSL pretraining (W2V2EN-AASIST) achieves stronger monolingual performance and moderate cross-lingual effectiveness but retains asymmetric transfer biases between language pairs. Language-specific fine-tuning (W2V2(xx)-AASIST) further enhances monolingual results while showing limited improvement in cross-lingual scenarios. Conversely, LFCC-ECAPA-TDNN demonstrate stronger cross-lingual robustness with comparable monolingual performance, maintaining stable accuracy across both high- and low-resource language groups. These findings indicate that (1) SSL models heavily depend on pretraining language coverage and fine-tuning strategies, whereas (2) LFCC-based approaches inherently prioritize language-agnostic acoustic features, enabling more reliable generalization across linguistic and resource-diverse conditions.

\subsection{Unseen Languages Generalization Experiment (RQ6)}

As shown in Table 5, the leave-one-language-out evaluation demonstrates generalization capabilities to unseen languages. The results reveal that XLSR-AASIST exhibits notable performance degradation on unseen languages, highlighting limitations due to its reliance on language-specific pretraining data. Furthermore, a trade-off between local and global modeling is observed: LFCC-ResNet18 effectively preserves local phoneme boundaries but struggles with cross-lingual prosody modeling, whereas LFCC-ECAPA-TDNN aggregates multi-scale temporal features, capturing both local articulatory details and global prosodic features through hierarchical modeling.

\subsection{Unseen Speakers Generalization Experiment (RQ7)}

Finally, Table \ref{tab:speaker-effects} reports the seen/unseen speaker analysis following the leave-one-language out setup with pseudo-speaker labels, broken down according to the held-out language and the three models. The findings related to the six languages and the three models are in line with the previous analyses. As for the relative performance for seen/unseen speakers, unlike was hypothesized, no apparent trends are visible---the results are dependent both on the held-out language and the model. While this may suggest that the investigated models can be robust to speaker factors, the number of speakers is low and the speaker labels were derived through clustering process. Another future study with larger number speakers and known target speaker labels is needed to validate these preliminary findings.

\section{Conclusion}
In this work, we establish the first multilingual benchmark for speech deepfake source tracing, covering both monolingual and cross-lingual scenarios across six languages and two model categories (DSP- and SSL-based models). Furthermore, we first  explore the effects of unseen languages and speakers in ST tasks. Our findings reveal three key insights: First, in monolingual scenarios, SSL front-ends fine-tuned on language-specific data outperform both multilingual/English-only pretrained SSL front-ends and LFCC front-ends. Second, LFCC features combined with ResNet or ECAPA-TDNN backends demonstrate superior cross-lingual generalization. Third, while cross-lingual generalization is stronger within the same language family, significant performance variations persist across language pairs. 

We also explored the impact of speaker variability, finding no consistent performance gap between seen and unseen pseudo-speakers, though results fluctuated across held-out languages and model types. This indicates a potential robustness to speaker variation, yet also highlights current limitations, including reliance on unsupervised speaker clustering and the lack of ground-truth speaker labels. As a future direction, we encourage further validation using larger, labeled datasets to better understand the interplay between language, speaker, and model-specific factors in deepfake attribution tasks. Our benchmark aims to establish a foundation for this emerging field and promote further research into multilingual, speaker-aware, and model-specific audio deepfake forensics.

\bibliographystyle{IEEEtran}
\bibliography{mybib}

\end{document}